\documentclass[12pt,preprint2]{aastex}

\newcommand\gtsim{\mathrel{\lower0.6ex\hbox{$\buildrel {\textstyle >}
      \over {\scriptstyle \sim}$}}}
\newcommand\ltsim{\mathrel{\lower0.6ex\hbox{$\buildrel {\textstyle <}
      \over {\scriptstyle \sim}$}}}

\slugcomment{To appear in Ap.J.}
\shorttitle{3C~133 NIR Jet}
\shortauthors{Floyd et al.}

\begin{document}

\title{An optical-IR jet in 3C~133\altaffilmark{1}}

\author{David J. E. Floyd\altaffilmark{2}\email{floyd@stsci.edu}
Robert Laing\altaffilmark{3},
Marco Chiaberge\altaffilmark{2},
Eric Perlman\altaffilmark{4},
William Sparks\altaffilmark{2},
Duccio Macchetto\altaffilmark{2},
Juan Madrid\altaffilmark{2},
David Axon\altaffilmark{5},
Christopher P. O'Dea\altaffilmark{5},
Stefi Baum\altaffilmark{5},
Alice Quillen\altaffilmark{6},
George Miley\altaffilmark{7},
Alessandro Capetti\altaffilmark{8}}

\altaffiltext{1}{Based on observations with the NASA/ESA Hubble Space Telescope, obtained at the Space Telescope Science Institute, which is operated by the Association of Universities for Research in Astronomy, Inc. (AURA), under NASA contract NAS5-26555}
\altaffiltext{2}{Space Telescope Science Institute, 3700 San Martin Drive, Baltimore, MD 21218, U.S.A.}
\altaffiltext{3}{ESO, Karl-Schwarzschild-Stra\ss e 2, D-85748, Garching-bei-M\"{u}nchen, Germany}
\altaffiltext{4}{Joint Center for Astrophysics, Physics Dept. University of Maryland, Baltimore County, 1000 Hilltop Circle, Baltimore, MD 21250}
\altaffiltext{5}{Department   of  Physics,   Rochester   Institute  of Technology, 85 Lomb Memorial Drive, Rochester, NY 14623.}
\altaffiltext{6}{Department  of Physics  and Astronomy,  University of Rochester, Bausch \& Lomb Hall, P.O. Box 270171, 600 Wilson Boulevard, Rochester, NY 14627.}
\altaffiltext{7}{Leiden Observatory, P.O. Box 9513, NL-2300 RA Leiden, The Netherlands.}
\altaffiltext{8}{INAF--Osservatorio Astronomico di Torino, Strada Osservatorio
20, 10025 Pino Torinese, Italy.}

\begin{abstract}
We report the discovery of a new optical-IR synchrotron jet in the radio galaxy 3C~133 from our HST/NICMOS snapshot survey. The jet and eastern hotspot are well resolved, and visible at both optical and IR wavelengths. The IR jet follows the morphology of the inner part of the radio jet, with three distinct knots identified with features in the radio. The radio-IR SED's of the knots are examined, along with those of two more distant hotspots at the eastern extreme of the radio feature. The detected emission appears to be synchrotron, with peaks in the NIR for all except one case, which exhibits a power-law spectrum throughout. 
\end{abstract}


\section{Introduction}
Non-thermal radiation at optical-IR wavelengths is a rare phenomenon, occurring only in high energy situations such as radio jets close to the nuclei of active galaxies, and in recently exploded supernova remnants. The jet phenomenon itself is ubiquitous in radio galaxies at radio wavelengths~\citep{bridleperley84} where fast, light outflows are rendered detectable through synchrotron emission of relativistic particles moving through magnetic fields. Synchrotron emission occurs across a wide wavelength range, but the number of jets identified at higher energies remains small.
Jets are now understood as an integral part of the standard model of AGN~\citep{begelman+84} as a means of transporting energy from central engine to the lobes of radio galaxies. There is strong evidence that a central super-massive black-hole provides the ultimate energy source, but the mechanisms of energy transport, particle acceleration, and collimation of the jet material are poorly understood. 

Strong magnetic fields may help accelerate particles to the high energies required to produce optical-IR synchrotron emission. However these fields also shorten the already brief radiative timescale ($\tau\propto B^{-2}\ltsim 10^{3}$~yr), thus making low magnetic fields the optimal sites for the detection of optical-IR jets and hotspots~\citep{brunetti+03}.
The short timescales make optical-IR jets powerful observational tools in understanding the physics of acceleration. They are believed to require recent (re-) acceleration, since the energy-loss timescales are shorter than the light-travel time from the central engine. Note, however, that the effects of relativistic beaming and time dilation need to be taken into account:~\citet{heinzbegelman97} and~\citet{gopal-krishna+01} suggest that there is no requirement for reacceleration if this is done.

To obtain a better understanding of these physical conditions in jets we need to study both their morphologies and their spectral energy distributions (SED's) across a large spectral range.
Thanks to new discoveries by the {\em Hubble Space Telescope} (HST), the number of known extragalactic optical-IR jets has grown rapidly over the last decade~\footnote{See http://home.fnal.gov/~jester/optjets/ and references therein}, but the majority of discoveries are in a single broad band and at comparatively low redshifts ($z\ll0.1$) due to the inherent faintness of the sources. The most detailed multiwavelength work has been done on prominent nearby or bright AGN such as M~87 (discovered by~\citealt{curtis1918}, but see~\citealt{waterszepf05} and references therein) and 3C~273 (e.g.~\citealt{jester+05}). A optical-IR wavelengths, emission from jets is generally interpreted as an extension of the synchrotron flux (e.g.~\citealt{scarpaurry02,sparks+94}), while at shorter wavelengths, more exotic radiative mechanisms come into play (see~\citep{sambruna+04} and references therein).

Here we present imaging of a newly-discovered optical-IR jet in 3C~133 (0459$+$252) from our {\em Near-Infrared Camera and Imaging Spectrometer} (NICMOS) snapshot program. The HST snapshot observing mode allows a large number of targets to be efficiently observed at irregular intervals that fill the scheduling gaps between other accepted GO programs. We now have a large body of data, obtained in this manner for the low-redshift 3CR sources, that has led to the serendipitous discovery of many of the currently known optical jets, and that now allows for multi-wavelength studies of these jets and their host galaxies. However we note that for detailed analysis of the physical conditions, X-ray observations which probe particles with even shorter radiative lifetimes than the optical, are proving invaluable with the advent of XMM and Chandra~\citep{sambruna+04}.

3C~133 is a classic example of a one-sided FRII~\citep{fanaroffriley74} radio galaxy showing weaker Faraday depolarisation with increasing wavelength in the lobe containing the jet~\citep{laing88}. However, it has an unusual asymmetric radio morphology, with a distinct $\sim90^\circ$ bend, and large scale morphology that has been interpreted as evidence for a rotating radio axis~\citep{robson81}, and by~\citet{williamsgull85} as evidence for a flow deflected by an oblique shock.
This is the second new jet to be discovered by this NICMOS program, GO-10173 (see also 3C~401,~\citealt{chiaberge+05}), and brings to 30 the total number of confirmed optical-IR jets. 

We assume throughout a flat, $\Lambda$-dominated cosmology, with $H_{0}=70$~km~s$^{-1}$Mpc$^{-1}$, $\Omega_{M}=0.3$. The optical counterpart of 3C~133 is at a redshift of $z=0.2775$.
At this distance, 1\arcsec corresponds to a linear scale of 4.217~kpc. The projected length of the radio jet is 15~kpc. The source lies at low Galactic latitude ($\lambda=-9^{\circ}.9$) and is thus heavily extinguished by our own Galaxy  (Galactic $A_{B}=4.096$~mag. -- \citealt{galextinct}). 

\section{Observations and Data Reduction}
\begin{figure*}[t]
\centering
\includegraphics[width=14cm]{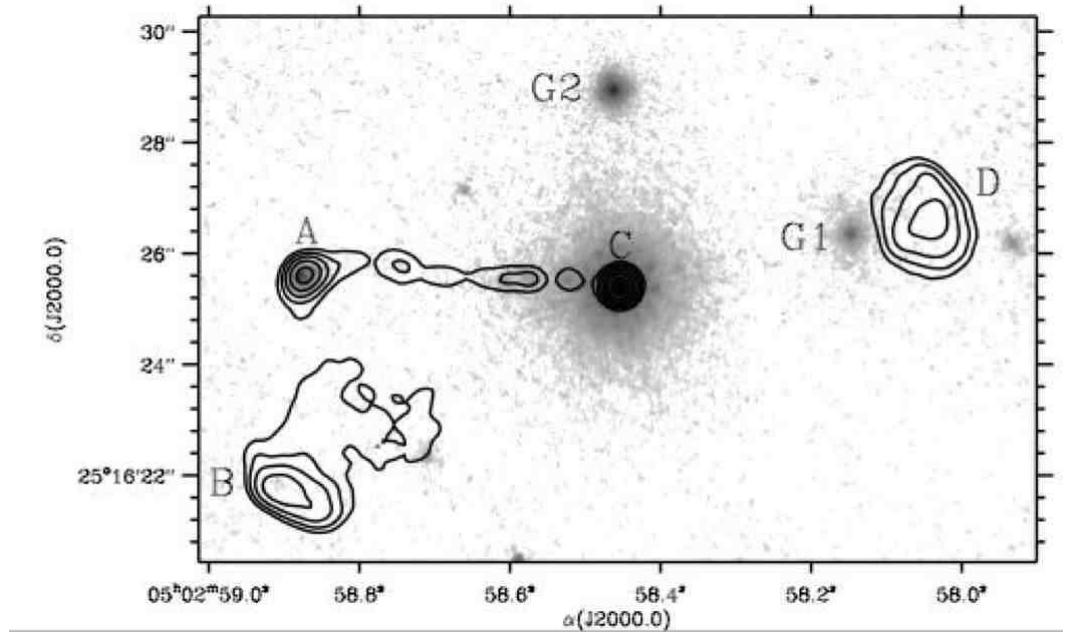} 
\caption{\label{fig-IR} The NIC2 F160W image overlayed with 6~cm radio contours (0\arcsec.35~FWHM). The four main radio features are labelled A, B, C, D, following the convention of Robson (1981). G1 and G2 indicate two nearby galaxies. Three distinct jet knots (E1, E2, E3) are observed east of the nucleus at this wavelength ($\lambda_{r}=1.25~\mu$m) -- See Fig. 2 for close-up detail.}
\end{figure*}

\subsection{Optical-IR observations}
Optical-IR observations are summarised in Table~\ref{tab-optobs}. 3C~133 was observed as part of our IR snapshot program (GO-10173) on 2004, December 13th, using NICMOS camera 2 (NIC2) and the F160W filter, which is centred at 1.6037~$\mu$m, and covers the spectral range from 1.4 to 1.8~$\mu$m. In the rest-frame of the source, the spectral range of the filter (1.1-1.4~$\mu$m) admits the Pa$\beta$ line at 1.28216~$\mu$m, but is broad enough that continuum emission dominates.
The field of view of NIC2 is $19\arcsec.2\times19\arcsec.2$, with a pixel size of $0\arcsec.076\times0\arcsec.075$. Total exposure time is 1152~s, split into 4 images of 288~s to allow for a 4-point dither pattern with sub-pixel spacing, in order to improve spatial resolution and eliminate detector artefacts and cosmic rays from the data. 
After dithering using the {\sc drizzle} algorithm of~\citet{drizzle} with a {\sc scale} factor of 0.5, we produce the final image presented in Fig.~\ref{fig-IR}, rotated to show north up, and with a projected pixel size of $0\arcsec.038\times0\arcsec.038$. 
The HST world coordinate system was corrected by comparison with the radio maps, using the galaxy's core as described below, to bring it onto the international coordinate reference frame (ICRF).
See Madrid et al. (2005 {\em in press}) for further details of the observations and data reduction. 

In order to obtain a measurement of the flux from the jet over as wide an energy range as possible, we examined archival optical data for the source. 3C~133 has been included in two previous HST/WFPC2 snapshot programs (Table~\ref{tab-optobs}). The HST/WFPC2 data were reduced as described in~\citet{dekoff+96}, and the sky background subtracted using a sigma-clipping technique. Note that the F702W observations were not CR-SPLIT, and no attempt was made at removing cosmic rays.

\begin{deluxetable}{rlrlc}
  \tabletypesize{\small}
  \tablecolumns{5}
  \tablewidth{0pc}
  \tablecaption{\label{tab-optobs}Journal of HST observations}
  \tablehead{
    \colhead{Filter} & \colhead{Camera} &
    \colhead{$t$/s} & \colhead{Obs. Date} &
    \colhead{ID}}
  \startdata
  \sc{f555w}  & WFPC2/PC &  300 & 1997 Sep 23 & 6967\\
  \sc{f702w}  & WFPC2/PC &  300 & 1994 Mar 12 & 5476$^\mathrm{a}$\\
  \sc{f160w}  & NIC2  & 1152 & 2004 Dec 13 & 10173$^\mathrm{b}$\\
  \enddata
  \tablecomments{HST optical-IR observations, with HST proposal ID's 
    and references where data is previously published.}
  \tablerefs{$^\mathrm{a}$ De Koff et al. (1996); 
    $^\mathrm{b}$ Madrid et al. (2005).}
\end{deluxetable}

\subsection{Radio observations}
3C~133 has been the subject of several VLA and MERLIN studies. We re-analysed archival data in order to produce four radio maps to compare to the IR data. Radio observations are summarised in Table~\ref{tab-radobs}. Each uv dataset was first calibrated and imaged separately.  The VLA~\footnote{The VLA is operated by the National Radio Astronomy Observatory. The National Radio Astronomy Observatory is a facility of the National Science Foundation operated under cooperative agreement by Associated universities, Inc.}
observations were reduced in the {\sc aips} package using standard techniques of amplitude and phase calibration, followed by imaging, {\sc clean} de-convolution and several iterations of self-calibration. The MERLIN~\footnote{MERLIN is a national facility operated by the University of Manchester on behalf of PPARC in the UK.}
data were also imaged and self-calibrated in {\sc aips} after initial calibration as described in~\citet{robson81}. The uv datasets for each of the three frequency bands were then concatenated, imaged and self-calibrated in order to improve the spatial coverage. The core varied by significant amounts between observations and this effect was corrected by imaging the individual datasets at the same resolution, measuring the maxima and adding or subtracting point sources as appropriate before concatenation. Final CLEAN images were made at resolutions of 0\arcsec.35 FWHM (2, 6, and 18~cm) and 0\arcsec.1 (2~cm only). Noise levels are given in Table~\ref{tab-radimages}.

\begin{deluxetable}{rrlrl}
  \tabletypesize{\small}
  \tablecolumns{5}
  \tablewidth{0pc}
  \tablecaption{\label{tab-radobs} Journal of radio observations}
  \tablehead{
    \colhead{$\nu$/GHz} & \colhead{$\Delta\nu$/GHz} &
    \colhead{Config.} & \colhead{$t/$min} & \colhead{Date}}
  \startdata
  14.9399 & 100 & VLA A  &  12 & 1985 Feb 18\\
  14.9649 &  50 & VLA B  &  26 & 1982 Aug 06\\
  14.9649 &  50 & VLA C  &  26 & 1983 May 05\\
   4.8851 &  50 & VLA A  &  72 & 1982 Mar 02 $^\mathrm{a}$\\
   1.6660 &   8 & MERLIN & 780 & 1980 Aug    $^\mathrm{b}$\\
   1.6649 &  50 & VLA A  &   8 & 1982 Mar 02\\
  \enddata
  \tablecomments{Radio observations, with references where data is previously published.}
  \tablerefs{$^\mathrm{a}$ Laing (1988); $^\mathrm{b}$ Robson (1981).}
\end{deluxetable}

\begin{deluxetable}{rrr}
  \tabletypesize{\small}
  \tablecolumns{3}
  \tablewidth{0pc}
  \tablecaption{\label{tab-radimages}Parameters of radio images}
  \tablehead{
    \colhead{$\nu$/GHz} & \colhead{Beam FWHM/arcsec} &
    \colhead{Noise/mJy~bm$^{-1}$}}
  \startdata
  14.9 & 0.1  & 0.06\\
  14.9 & 0.35 & 0.50\\		     
  4.9  & 0.35 & 0.07\\
  1.7  & 0.35 & 0.39\\
\enddata
\tablecomments{Radio image beam size and noise.}
\end{deluxetable}

\section{Results}
\label{sec-results}
\subsection{Discovery of a new jet}
3C~133 is found to exhibit a prominent NIR jet (Fig.~\ref{fig-IR}). In addition to the host galaxy, three distinct bright sources are clearly visible (A, G1, G2) as well as a number of fainter sources, and a line of three faint ``knots'' comprising  the jet, extending $\sim 1-2$~arcsec east of the nucleus along the axis of the radio source. \citet{dekoff+96} noted the existence of faint optical emission extending east, positing it as an optical jet candidate. 

\begin{figure*}[t]
\centering
\includegraphics[width=8cm]{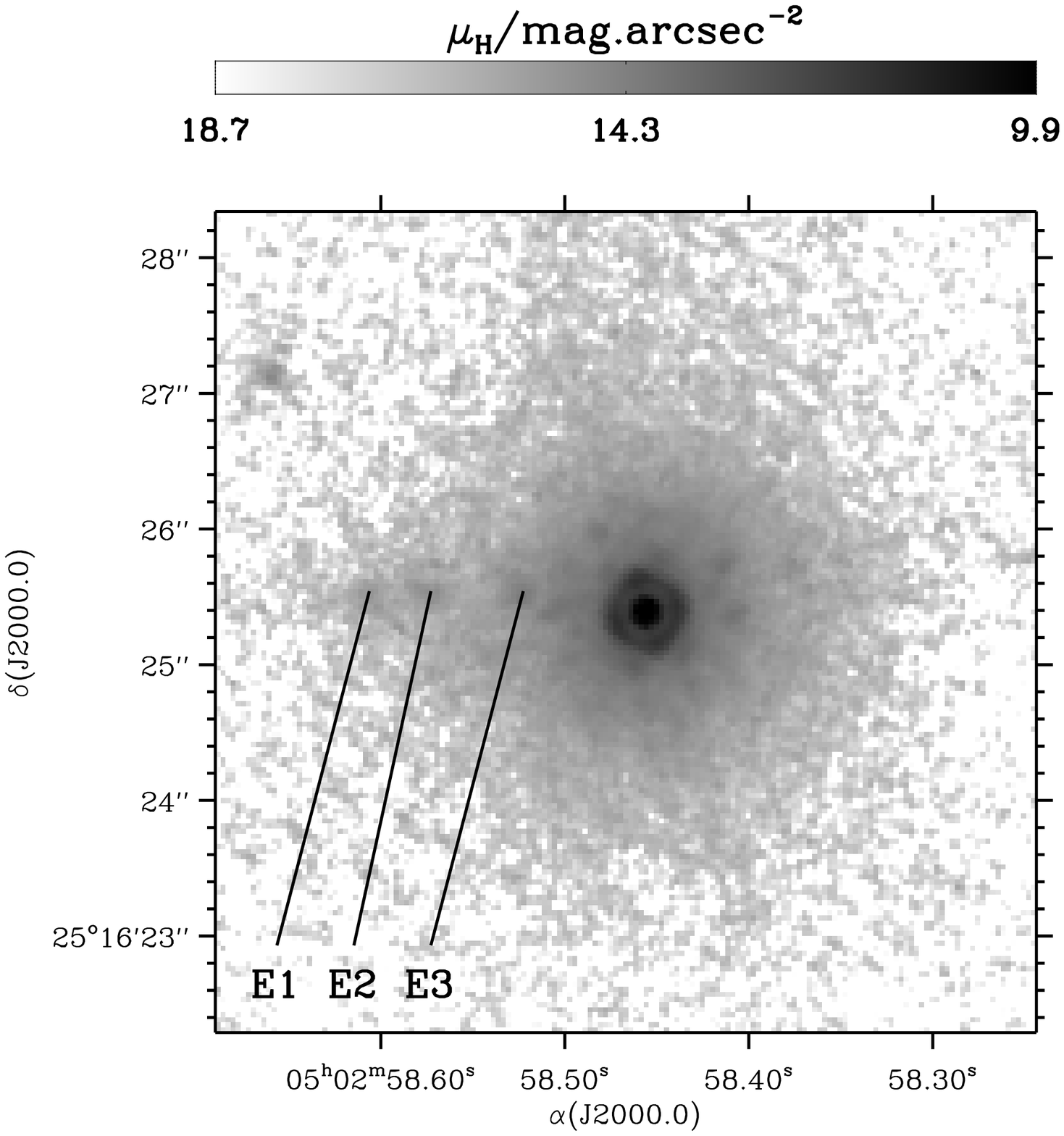}
\includegraphics[width=8cm]{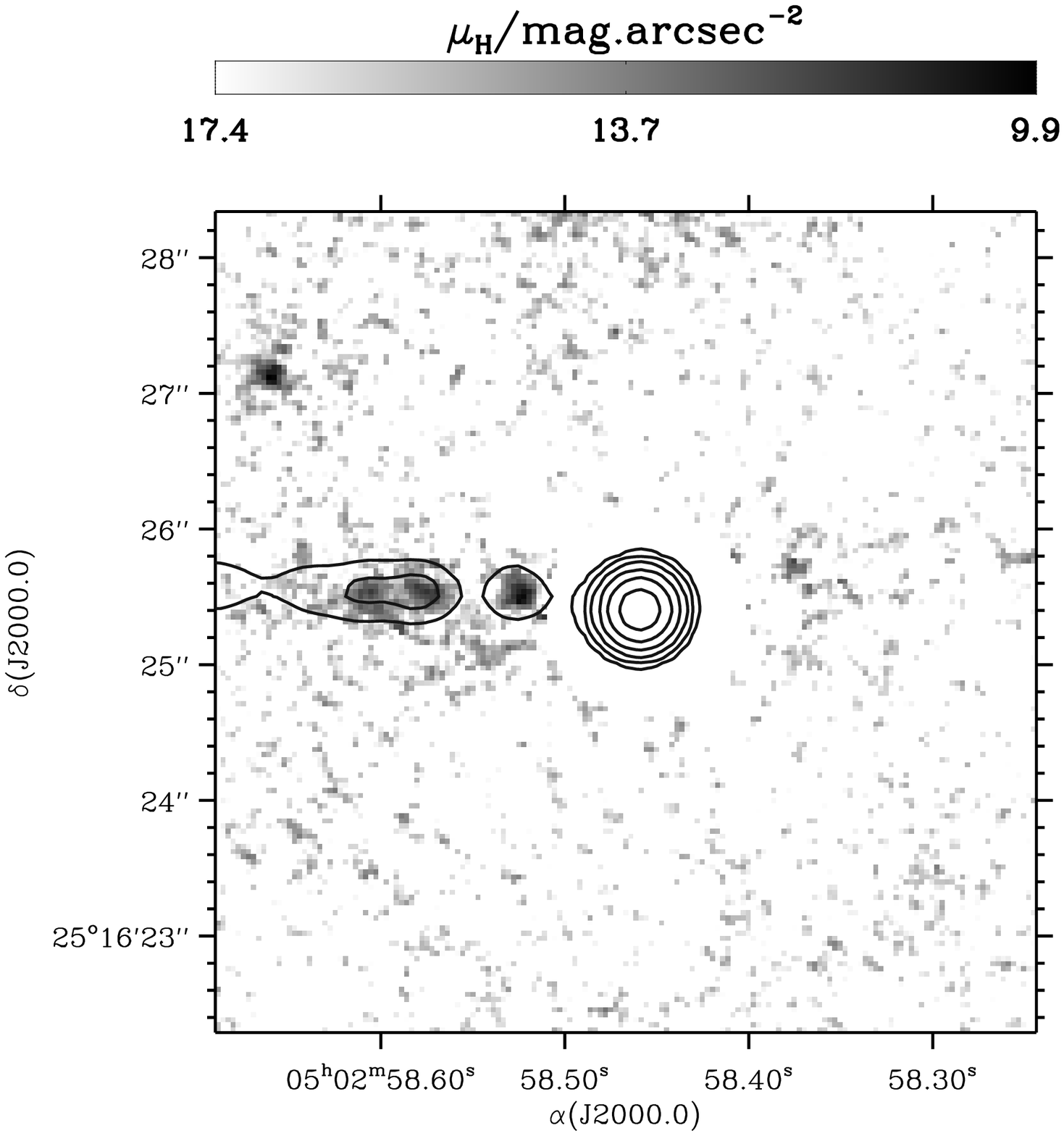}
\caption{\label{fig-IR-eres} Close up of jet detail (left), with the nucleus masked in the ellipse model-subtracted NIC2 F160W image (right) overlayed with 6~cm radio contours (0\arcsec.35~FWHM). PSF artefacts are visible close to the masked region, out to 0\arcsec.76 radius.}
\end{figure*}

We modelled the IR emission from the central IR source using an isophotal {\sc ellipse} fit in {\sc iraf}. Obvious external sources, including the jet itself were masked from the fit, as was the central 20 pixel (0\arcsec.76) radius where the asymmetries of the instrumental PSF dominate the flux. The ellipse model-subtracted residuals (Fig.~\ref{fig-IR-eres}) were convolved with a Gaussian function and interpolated onto the grid of the 2~cm map to match the beam size, resolution and astrometry of the radio data. Similar modelling was performed for the 2 optical images. The residual images were used to perform photometry on the jet itself (section~\ref{sec-charjet} below).

\begin{figure*}[h]
\centering
\includegraphics[width=8cm]{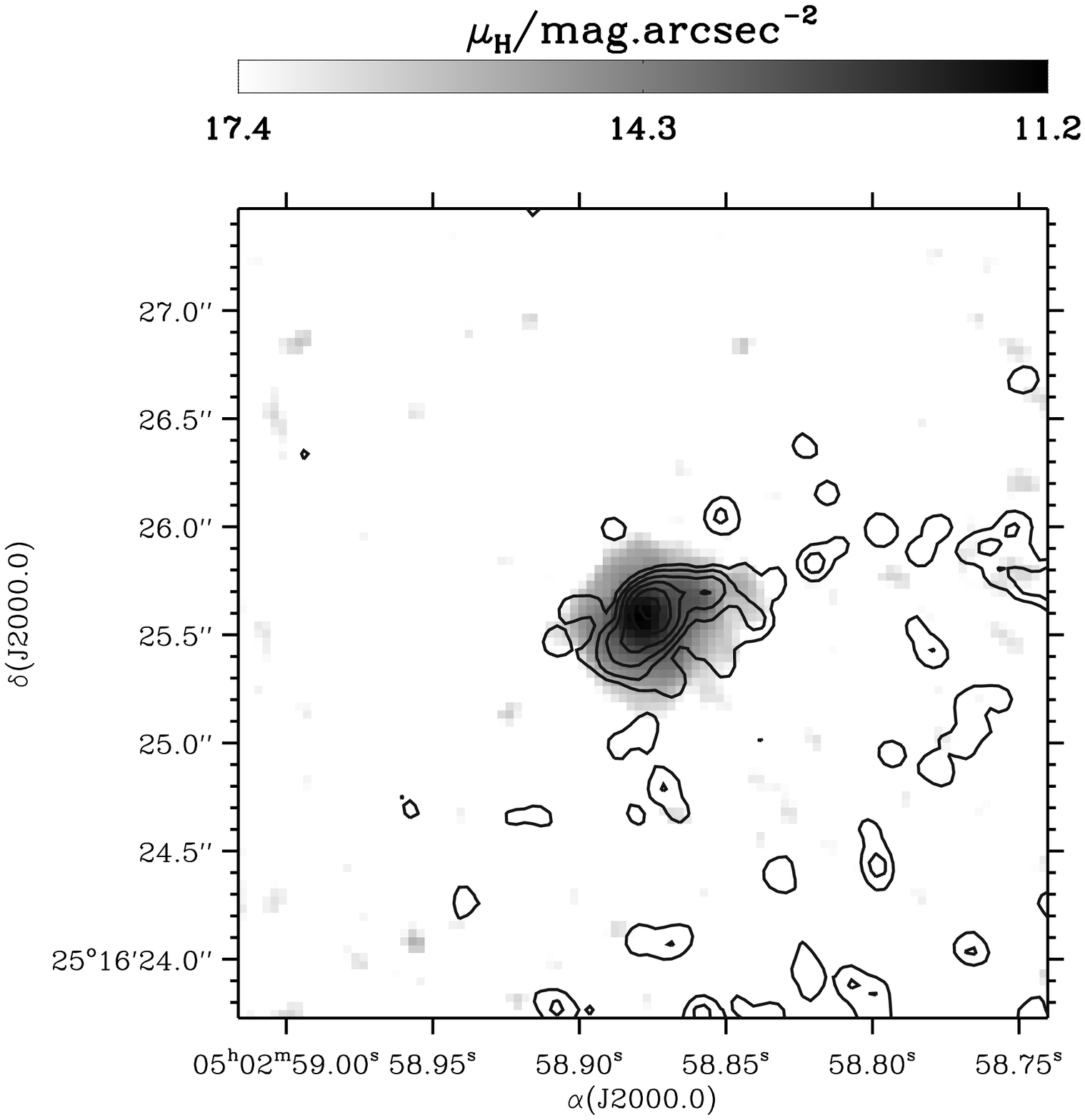} 
\includegraphics[width=8cm]{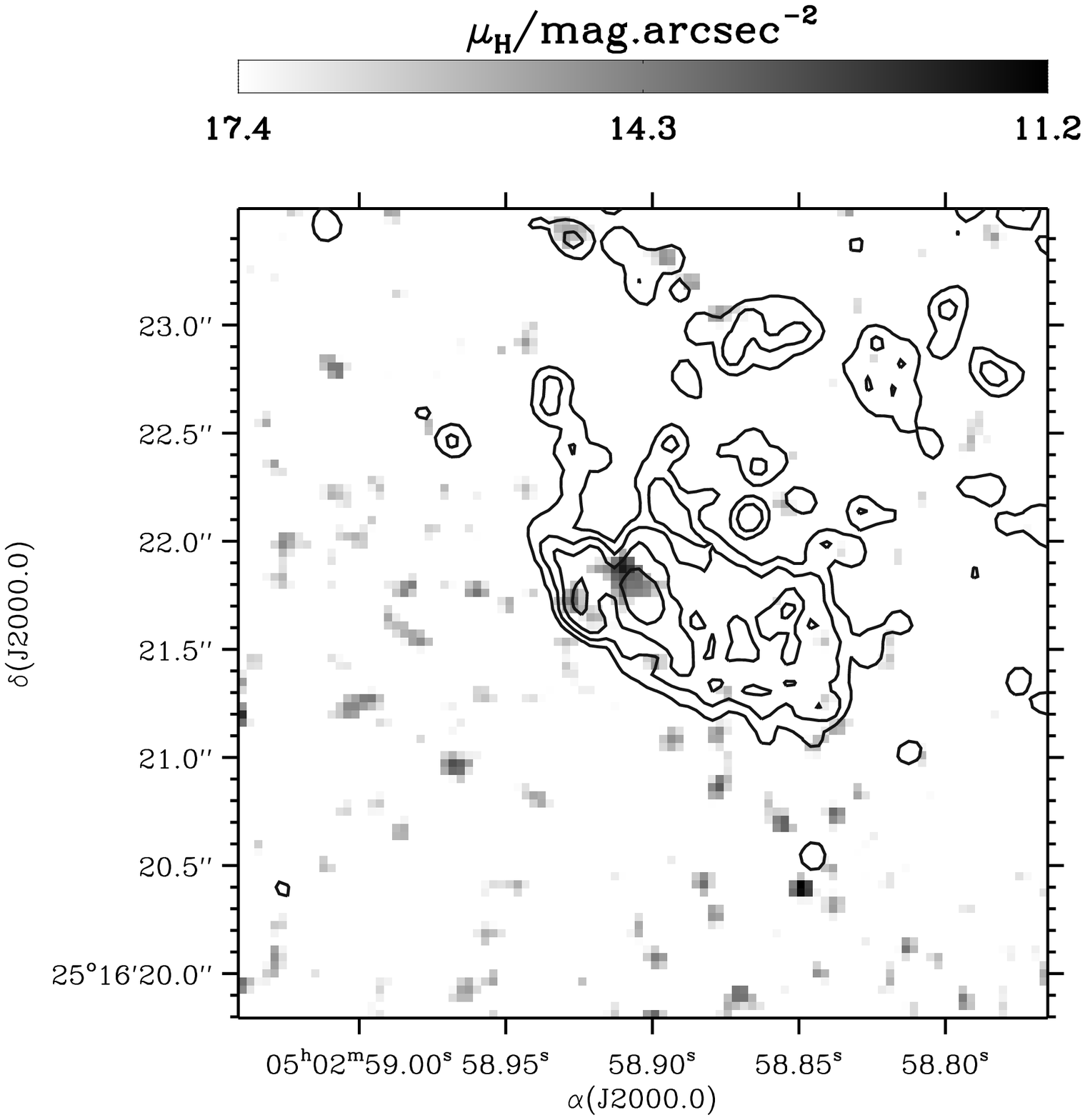} 
\caption{\label{fig-AB} The smoothed NICMOS2 F160W image showing knots A (left) and tentative, 4$\sigma$, detection of B (right), overlayed with 2~cm radio contours (0\arcsec.1~FWHM).}
\end{figure*}

The three IR knots clearly align with three of the brightest (6~cm) radio features in the jet, which we herein refer to as E1--E3, moving from east to west towards the nucleus. Furthermore, the eastern IR feature aligns with Robson's (1981) component A (Fig.~\ref{fig-AB}), and the nucleus with component C. To the south-east, component B is faintly detected (4$\sigma$) in the IR at a position close to the peak of the radio flux (Fig.~\ref{fig-AB}). Faint, low-significance detections are made in the F555W and F702W images at the same locations, although in the F702W image, B is partially obscured by a cosmic ray hit (as the observations were not CR-SPLIT). 
We also detect a string of faint (2--3$\sigma$) features on the western side, up to 2\arcsec.8 away from the nucleus, beyond the PSF artefacts. This is in direct line between the core (radio component C) and radio component D, but no significant radio flux is detected on the counter-jet side at 6 or 18~cm. 


\subsection{Characterising the jet}
\label{sec-charjet}
Photometry was performed in five box-shaped apertures on each image, corresponding to the positions of knots E1, E2, E3 and the jet hotspots A, B. The aperture sizes and detected fluxes are reported in Table~\ref{tab-SED}. 
For each feature, identical apertures were used at each wavelength. In the optical-IR, the Gaussian-smoothed, rebinned ellipse model-subtracted residual images (discussed above) were used for the photometry, and they match the beam size and resolution of the 2~cm image. 
The mean error on the ellipse fit flux at the radius of each feature was added in quadrature to the Poisson noise of the feature itself, although this is a small contribution to the noise for all except E3.
For the two lower frequency radio maps (6 \&18~cm), fluxes were measured at the lower 0\arcsec.35 resolution. 
Aperture corrections were calculated for each feature by comparing to results obtained by applying the same apertures to a {\sc tinytim}~\citep{tinytim} realization of a point source for the HST images, and a Gaussian of the appropriate beam size for the radio images. We note that hotspot B exhibits a very different morphology in the radio to the infrared (Fig.~\ref{fig-AB}). A range of aperture sizes were tested, and used to compute the error on the flux from hotspot B.

Galactic extinction was corrected for following~\citet{schlegel+98}.
There may be an additional extinction contribution from the host galaxy itself, but this is not computed here due to the poor SNR on the host galaxy colour measurements. Clearly any such contribution will affect the innermost feature E3 at the shortest wavelengths the most, and would likely steepen the observed SED.

\begin{figure*}[ht]
\centering
\includegraphics[height=15cm,angle=90]{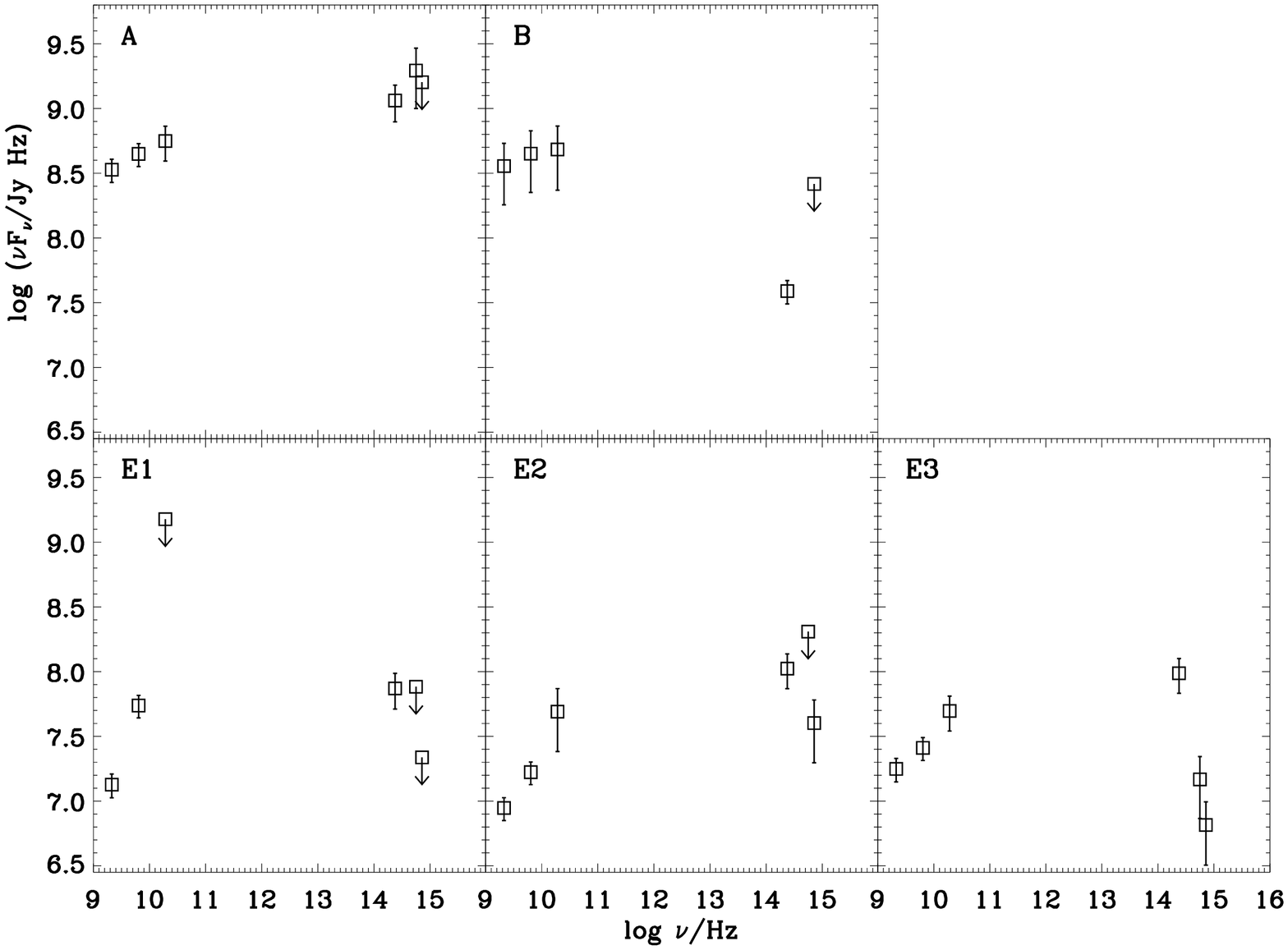} 
\caption{\label{fig-SED} SED's for the 3 jet components (E1, E2, E3) and the hotspots (A, B). All except A are consistent with a synchrotron spectrum with a turnover at $\sim10^{13}$~Hz. See Table~\ref{tab-SED} and discussion.}
\end{figure*}

We used the NIR and 6~cm data to determine the radio-IR spectral index, $\alpha$ ($F_{\nu}\propto\nu^{-\alpha}$), of each of the five major features (see Table~\ref{tab-SED}). These spectral indices are almost equal in E2, E3, and A ($\alpha\approx0.9$), with E1 and B having somewhat steeper spectra ($\alpha_{E1}=0.96\pm0.03;~\alpha_{B}=1.17\pm0.03$). The spectral energy distributions (SED's) of the six components are shown in Fig.~\ref{fig-SED}. 

\subsection{The host galaxy and source G1}
We modelled the IR source using the technique described in~\cite{floyd+04} to separate host from nuclear flux. The host galaxy is found to be well fit by a de Vaucouleurs profile with $R_{e}=2$~kpc and a total unobscured $H$-band magnitude of 16.14. 
The unresolved nucleus has $H=17.06$, contributing roughly one third of the total flux at this wavelength. The optical host is harder to separate from the nucleus, due to the lower SNR of the images. We determined an R-band host magnitude of $18.6\pm1.0$, giving a colour of $R-H=2.5\pm1.0$, consistent with that of an old stellar population ($R-H\approx2.3$).

Although the western feature on the IR image, G1, sits snugly alongside the exterior contours of the western radio hotspot, D, its colours are very red ($R-H\approx2.6\pm0.5$) and it is resolved in the IR. 
It thus seems likely to be a small companion or background elliptical galaxy with a mature stellar population.

\begin{deluxetable}{rllllllll}
  \tabletypesize{\tiny}
  \tablecolumns{9}
  \tablewidth{0pc}
  \tablecaption{\label{tab-SED}Jet SED}
  \tablehead{
    \colhead{Src} & \colhead{$l_{AP}$} &
    \colhead{$F_\nu(7.18E14\mathrm{Hz})$}&\colhead{$F_\nu(5.62E14\mathrm{Hz})$}&
    \colhead{$F_\nu(2.38E14\mathrm{Hz})$}&\colhead{$F_\nu(1.91E10\mathrm{Hz})$}&
    \colhead{$F_\nu(6.38E9 \mathrm{Hz})$}&\colhead{$F_\nu(2.13E9\mathrm{Hz})$}&
    \colhead{$\alpha_{\mathrm radio-IR}$}\\
    \colhead{} & \colhead{} &
    \colhead{Jy}&\colhead{Jy}&
    \colhead{Jy}&\colhead{mJy}&
    \colhead{mJy}&\colhead{mJy}&
    \colhead{}}
  \startdata
  E1 & 0\arcsec.3 & $<3.03E-8       $ & $<1.36E-7       $ & $ 3.12\pm0.94E-7 $ & $<7.86       $ & $ 8.58\pm1.7$  & $ 6.30\pm1.3 $ & $0.96\pm0.03$ \\ 
  E2 & 0\arcsec.4 & $ 5.59\pm2.8E-8 $ & $<3.63E-7       $ & $ 4.43\pm1.3E-7  $ & $ 2.56\pm1.3 $ & $ 2.62\pm0.52$ & $ 4.16\pm0.83$ & $0.87\pm0.03$ \\ 
  E3 & 0\arcsec.4 & $ 9.12\pm4.6E-9 $ & $ 2.62\pm1.3E-8 $ & $ 4.08\pm1.2E-7  $ & $ 2.60\pm0.78$ & $ 4.04\pm0.81$ & $ 8.33\pm1.7 $ & $0.88\pm0.03$ \\ 
  A  & 1\arcsec.1 & $<2.22E-6       $ & $ 3.50\pm1.7E-6 $ & $ 4.84\pm1.5E-6  $ & $ 29.3\pm8.8 $ & $ 69.8\pm14.0$ & $ 158.\pm32.0$ & $0.91\pm0.03$ \\ 
  B  & 0\arcsec.8 & $<3.64E-7$        & $<1.63E-6       $ & $ 1.63\pm0.32E-7 $ & $ 25.2\pm13.0 $ & $ 70.2\pm35.0$ & $ 169.\pm84.0$ & $1.17\pm0.13$ \\ 
  \enddata
  \tablecomments{Extinction-corrected optical, IR, and radio fluxes of the five synchrotron sources in 3C~133 at the rest-frame frequency, $\nu$ of the observations. 3$\sigma$ upper limits are given in the event of a non-detection. 6~cm--1.6~$\mu$m spectral indices ($F_{\nu}\propto\nu^{-\alpha}$) are given in the final column. Apertures are square boxes with the length of one side presented in column 2.}
\end{deluxetable}

\begin{figure}[ht]
\centering
\includegraphics[height=8cm]{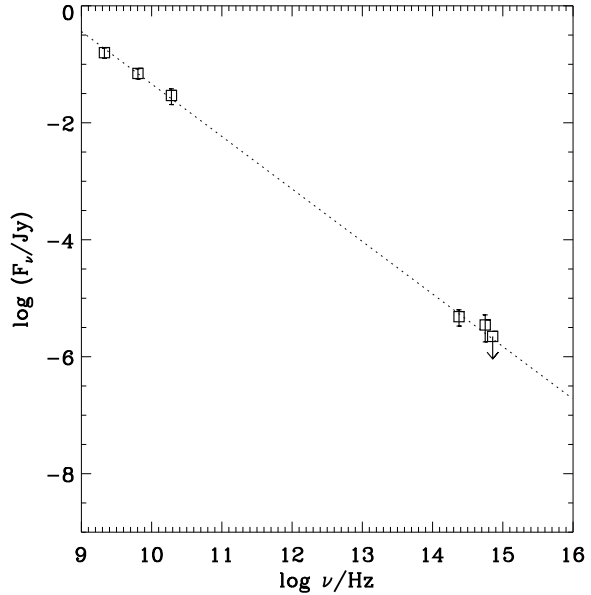} 
\caption{\label{fig-SEDA} SED for hotspot A, showing the best linear fit to the data between 6~cm and 0.7~$\mu$m (dotted line; $\log_{10}(F_\nu/Jy)=7.64-0.90\log_{10}(\nu/Hz)$). This fit gives a flux of $2.5 \times 10^{-6}$ Jy at 0.55~$\mu$m, just within the 3$\sigma$ upper limit on the F555W datum.}
\end{figure}

\section{Discussion}
\label{sec-disc}
\subsection{Synchrotron Jet}
For all regions except A the knots are consistent with synchrotron emission in the optical with $\nu_\mathrm{peak}\sim10^{11}-10^{15}$~Hz in $\nu F_\nu$. 
Component A itself has a spectrum consistent with a single power-law ($\alpha=0.90\pm0.02;~\chi^2=1.2$ with 3 degrees of freedom) between 6~cm and 0.7~$\mu$m, although the spectrum appears to flatten within the optical-IR band (Fig.~\ref{fig-SED}). The flux density predicted by this fit at 0.55~$\mu$m is $2.5 \times 10^{-6}$ Jy, just within the $3\sigma$ upper limit.
Synchrotron emission from a single population is the most likely explanation of the spectrum, but we note that if this is the case then the SED turnover frequency $\nu_\mathrm{peak}>10^{15}$~Hz, is unusually high but not unprecedented~\citep{brunetti+03}. The hint of a flattening in the spectrum at high frequencies, and the extreme value of the turnover frequency suggest that a second component may contribute to the flux at high frequencies. This could be synchrotron from a second population, or inverse Compton scattering of CMB photons by very low-$\gamma$ electrons~\citep{georganopoulos+05}. 
In itself, it is unsurprising that the SED of A differs from that of the knots in the jet; The X-ray emission from hotspots has been well studied (e.g.~\citealt{hardcastle+04,brunetti+03,hardcastle+02,wilson+01}) and always found to be quite hard. However, it is impossible to distinguish between the different scenarios without high-resolution ultraviolet or X-ray imaging.

We note the apparent decrease in the turnover energy from E3 to E1, which accompanies the steepening radio-IR spectral index. The peak energy drops with distance from the nucleus, and at E1 appears to be extremely low. The optical-radio spectral indices of the jet knots in 3C~133 are lower than those for the majority of the jets studied by~\citet{sambruna+04}. X-ray observations of 3C~133 would therefore allow us to study an undersampled part of the $\alpha$ (radio -- optical) -- $\alpha$ (optical -- X-ray) plane and to distinguish between synchrotron and beamed inverse Compton as the primary X-ray emission mechanism.

\subsection{Jet Orientation}
We note that the jet is strongly one-sided, implying that the eastern jet is approaching and relativistically beamed. The nuclear point source is luminous ($M_{H}(\mathrm{Nuc})=-23.71$) and almost quasar-like, according to the definition of~\citealt{VCV03}. This suggests that we are seeing the source at an inclination angle $<45^{\circ}$, if we adopt the standard model picture~\citep{barthel89,urrypadovani95,jacksonwall95}. 3C~133 is known to exhibit an emission line spectrum~\citep{smithspinrad80} with prominent [O~{\sc iii}] as well as H$\beta$ emission.
The presence of typically broad lines that are seen in QSO's would support a picture of an object that is being observed close to pole-on. 

\begin{figure*}[ht]
\centering
\includegraphics[width=8cm]{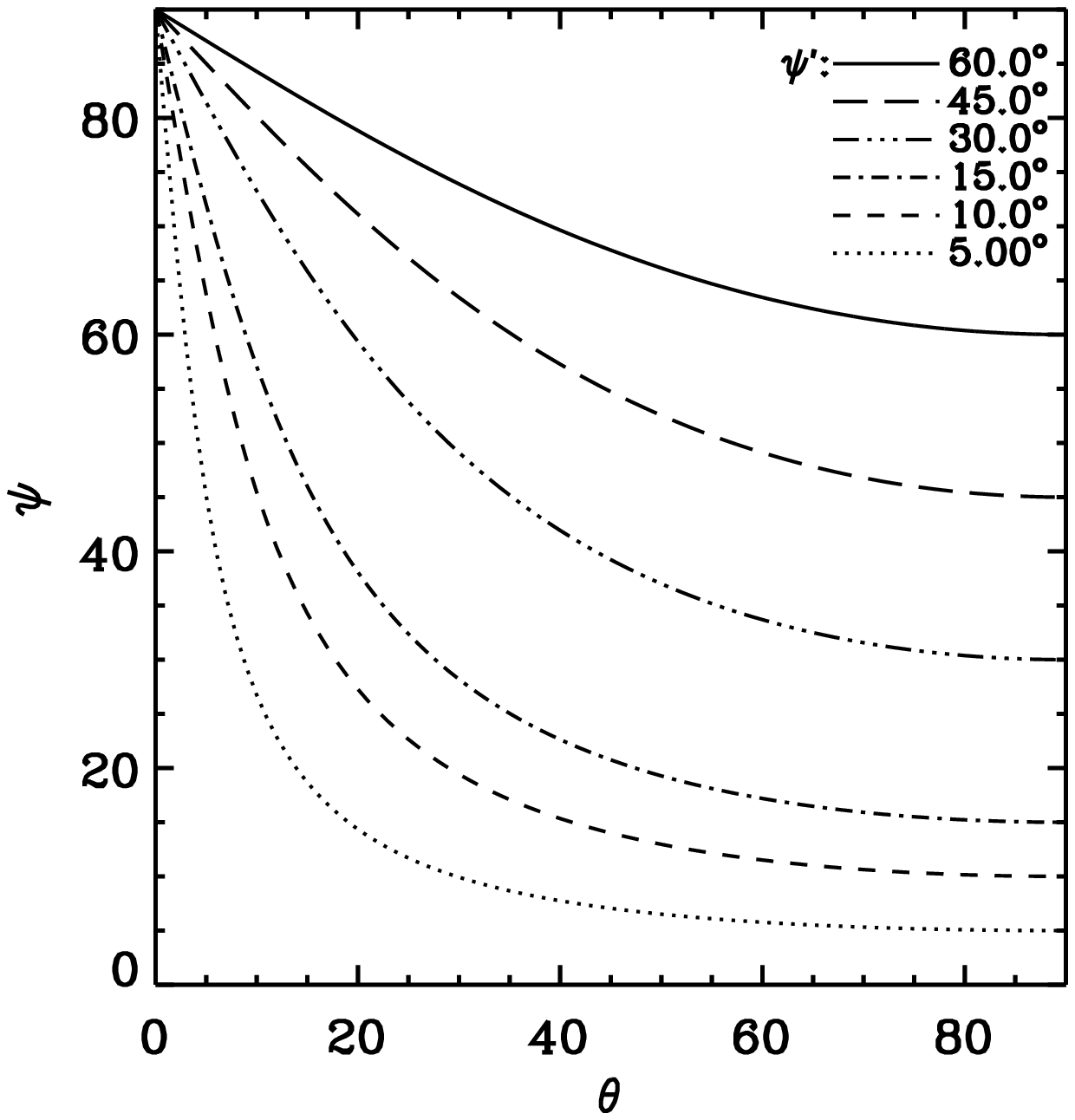} 
\includegraphics[width=8cm]{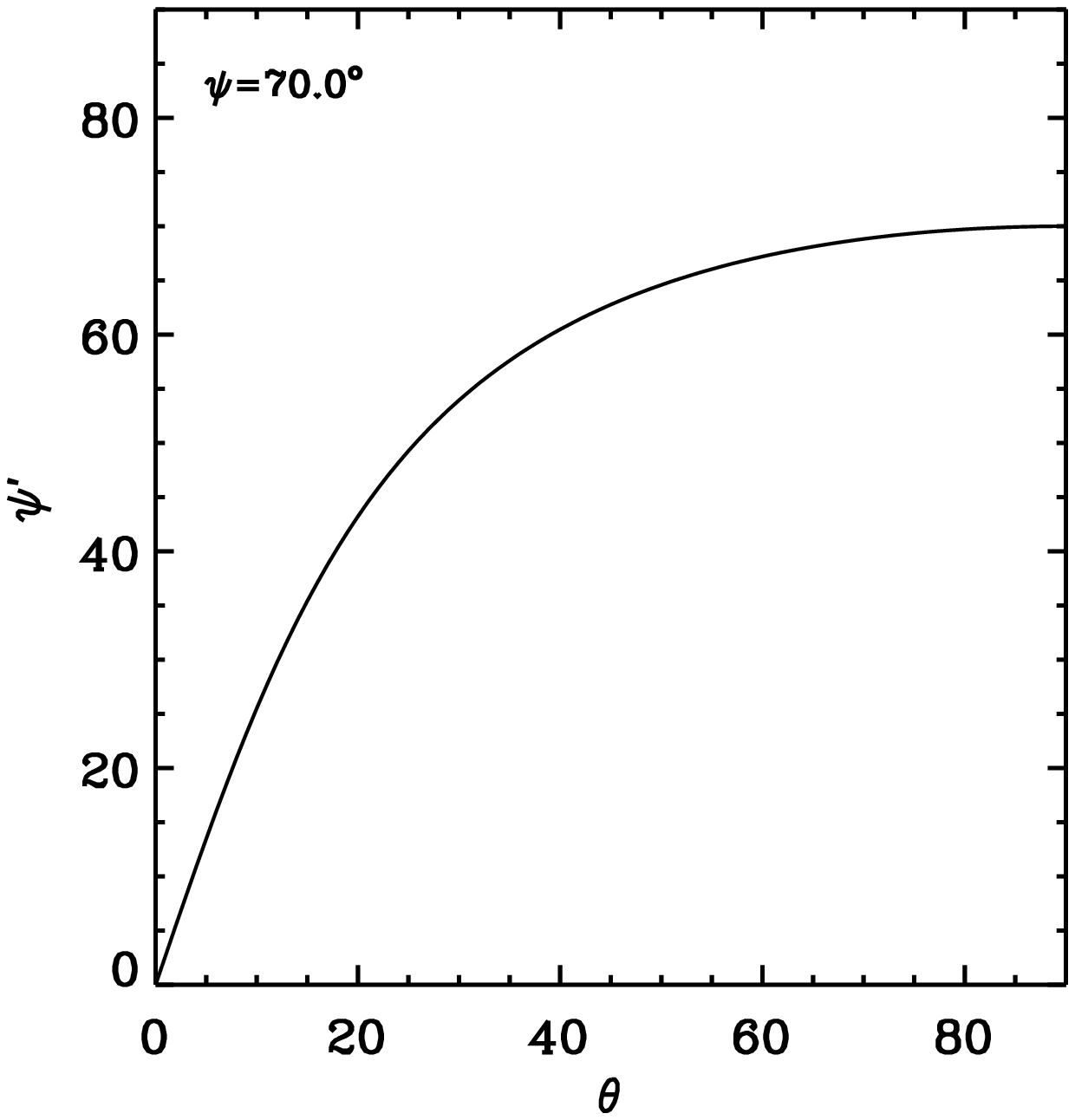} 
\caption{\label{fig-psi} True ($\psi^\prime$) and apparent ($\psi$) bend angles, based on orientation angle, $\theta$.}
\end{figure*}

The jet has an apparent bend of $\approx70^\circ$ at A, where the jet flow appears to be deflected, perhaps by an oblique shock~\citep{williamsgull85}, and ceases to be a collimated jet-like structure. We interpret this as a much smaller bend in a jet that is oriented close to the line of sight. In Fig.~\ref{fig-psi} we show the apparent bend angle on the sky, $\psi$, for different values of true bend angle, $\psi^\prime$, and orientation angle, $\theta$ ($\tan\psi\propto1/\sin\theta$). We also plot the true bend angle for the case at A, $\psi=70^\circ$. This shows that unless the true bend angle, $\psi^\prime$ is $>60^\circ$, the jet must be oriented closer to the line of sight than ($\theta\ltsim40^\circ$), confirming and strengthening the suggestion above. Finally, the rest-frame core-dominance ratio at 2~cm, $R=P_\mathrm{core}/P_\mathrm{ext} =0.2$ is in the top tenth percentile of the distribution derived by~\citet{hardcastle+99} and the core flux is $\approx17$ times larger than the median value expected from the~\citet{giovannini+01} correlation between total and core power. This high core fraction is also consistent with a small angle to the line of sight.

\section{Conclusions}
The IR and faint optical flux to the east of the nucleus appears to be synchrotron-dominated radiation from the jet. There is a slight steepening of the radio-IR spectral index as we move outwards as observed in, for example 3C~401~\citep{chiaberge+05}, 3C~293 (Floyd et al. {\em in press}), and 3C~273~\citep{jester+05}. 
However, high-resolution X-ray observations are required in order to properly diagnose the dominant radiation mechanism, and physical conditions inside the jet. We propose that hotspot A is at the working surface of the jet, and is the location of a significant bend, a likely site for in situ acceleration (similar to the situation in 3C~351~\citealt{hardcastle+02}). The jet is aligned close to the line of sight, and this deflection while small in itself (perhaps $\sim10^\circ$) produces the appearance of a $70^\circ$ kink in the jet on the plane of the sky. 

\section*{}
We gratefully acknowledge support from HST grant STGO-10173.
EP acknowledges support from NASA LTSA grants NAG5-9997 and NNG05-GD63G. We also thank the anonymous referee for constructive comments and suggestions in the preparation of this paper.

\bibliographystyle{astron}
\bibliography{floyd_etal_3c133}

\begin{thebibliography}{}

\bibitem[\protect\astroncite{{Barthel}}{1989}]{barthel89}
{Barthel}, P.~D.: 1989,
\newblock {\em \apj} {\bf 336}, 606

\bibitem[\protect\astroncite{{Begelman} et~al.}{1984}]{begelman+84}
{Begelman}, M.~C., {Blandford}, R.~D., and {Rees}, M.~J.: 1984,
\newblock {\em Reviews of Modern Physics} {\bf 56}, 255

\bibitem[\protect\astroncite{{Bridle} and {Perley}}{1984}]{bridleperley84}
{Bridle}, A.~H. and {Perley}, R.~A.: 1984,
\newblock {\em \araa} {\bf 22}, 319

\bibitem[\protect\astroncite{{Brunetti} et~al.}{2003}]{brunetti+03}
{Brunetti}, G., {Mack}, K.-H., {Prieto}, M.~A., and {Varano}, S.: 2003,
\newblock {\em \mnras} {\bf 345}, L40

\bibitem[\protect\astroncite{{Chiaberge} et~al.}{2005}]{chiaberge+05}
{Chiaberge}, M., {Sparks}, W.~B., {Macchetto}, F.~D., {Perlman}, E., {Capetti},
  A., {Balmaverde}, B., {Floyd}, D., {O'Dea}, C., and {Axon}, D.~J.: 2005,
\newblock {\em \apj} {\bf 629}, 100

\bibitem[\protect\astroncite{{Curtis}}{1918}]{curtis1918}
{Curtis}, H.~D.: 1918,
\newblock {\em Publications of Lick Observatory} {\bf 13}, 0

\bibitem[\protect\astroncite{{de Koff} et~al.}{1996}]{dekoff+96}
{de Koff}, S., {Baum}, S.~A., {Sparks}, W.~B., {Biretta}, J., {Golombek}, D.,
  {Macchetto}, F., {McCarthy}, P., and {Miley}, G.~K.: 1996,
\newblock {\em \apjs} {\bf 107}, 621

\bibitem[\protect\astroncite{{Fanaroff} and {Riley}}{1974}]{fanaroffriley74}
{Fanaroff}, B.~L. and {Riley}, J.~M.: 1974,
\newblock {\em \mnras} {\bf 167}, 31P

\bibitem[\protect\astroncite{{Floyd} et~al.}{2004}]{floyd+04}
{Floyd}, D.~J.~E., {Kukula}, M.~J., {Dunlop}, J.~S., {McLure}, R.~J., {Miller},
  L., {Percival}, W.~J., {Baum}, S.~A., and {O'Dea}, C.~P.: 2004,
\newblock {\em \mnras} {\bf 355}, 196

\bibitem[\protect\astroncite{{Fruchter} and {Hook}}{2002}]{drizzle}
{Fruchter}, A.~S. and {Hook}, R.~N.: 2002,
\newblock {\em \pasp} {\bf 114}, 144

\bibitem[\protect\astroncite{{Georganopoulos} et~al.}{2005}]{georganopoulos+05}
{Georganopoulos}, M., {Kazanas}, D., {Perlman}, E., and {Stecker}, F.~W.: 2005,
\newblock {\em \apj} {\bf 625}, 656

\bibitem[\protect\astroncite{{Giovannini} et~al.}{2001}]{giovannini+01}
{Giovannini}, G., {Cotton}, W.~D., {Feretti}, L., {Lara}, L., and {Venturi},
  T.: 2001,
\newblock {\em \apj} {\bf 552}, 508

\bibitem[\protect\astroncite{{Gopal-Krishna} et~al.}{2001}]{gopal-krishna+01}
{Gopal-Krishna}, {Subramanian}, P., {Wiita}, P.~J., and {Becker}, P.~A.: 2001,
\newblock {\em \aap} {\bf 377}, 827

\bibitem[\protect\astroncite{{Hardcastle} et~al.}{1999}]{hardcastle+99}
{Hardcastle}, M.~J., {Alexander}, P., {Pooley}, G.~G., and {Riley}, J.~M.:
  1999,
\newblock {\em \mnras} {\bf 304}, 135

\bibitem[\protect\astroncite{{Hardcastle} et~al.}{2002}]{hardcastle+02}
{Hardcastle}, M.~J., {Birkinshaw}, M., {Cameron}, R.~A., {Harris}, D.~E.,
  {Looney}, L.~W., and {Worrall}, D.~M.: 2002,
\newblock {\em \apj} {\bf 581}, 948

\bibitem[\protect\astroncite{{Hardcastle} et~al.}{2004}]{hardcastle+04}
{Hardcastle}, M.~J., {Harris}, D.~E., {Worrall}, D.~M., and {Birkinshaw}, M.:
  2004,
\newblock {\em \apj} {\bf 612}, 729

\bibitem[\protect\astroncite{{Heinz} and {Begelman}}{1997}]{heinzbegelman97}
{Heinz}, S. and {Begelman}, M.~C.: 1997,
\newblock {\em \apj} {\bf 490}, 653

\bibitem[\protect\astroncite{{Jackson} and {Wall}}{1999}]{jacksonwall95}
{Jackson}, C.~A. and {Wall}, J.~V.: 1999,
\newblock {\em \mnras} {\bf 304}, 160

\bibitem[\protect\astroncite{{Jester} et~al.}{2005}]{jester+05}
{Jester}, S., {R{\" o}ser}, H.-J., {Meisenheimer}, K., and {Perley}, R.: 2005,
\newblock {\em \aap} {\bf 431}, 477

\bibitem[\protect\astroncite{{Krist}}{1999}]{tinytim}
{Krist}, J.: 1999,
\newblock {\em TinyTim User Manual}

\bibitem[\protect\astroncite{{Laing}}{1988}]{laing88}
{Laing}, R.~A.: 1988,
\newblock {\em \nat} {\bf 331}, 149

\bibitem[\protect\astroncite{{Robson}}{1981}]{robson81}
{Robson}, D.~W.: 1981,
\newblock {\em \nat} {\bf 294}, 57

\bibitem[\protect\astroncite{{Sambruna} et~al.}{2004}]{sambruna+04}
{Sambruna}, R.~M., {Gambill}, J.~K., {Maraschi}, L., {Tavecchio}, F.,
  {Cerutti}, R., {Cheung}, C.~C., {Urry}, C.~M., and {Chartas}, G.: 2004,
\newblock {\em \apj} {\bf 608}, 698

\bibitem[\protect\astroncite{{Scarpa} and {Urry}}{2002}]{scarpaurry02}
{Scarpa}, R. and {Urry}, C.~M.: 2002,
\newblock {\em New Astronomy Review} {\bf 46}, 405

\bibitem[\protect\astroncite{{Schlegel} et~al.}{1998a}]{galextinct}
{Schlegel}, D.~J., {Finkbeiner}, D.~P., and {Davis}, M.: 1998a,
\newblock {\em \apj} {\bf 500}, 525

\bibitem[\protect\astroncite{{Schlegel} et~al.}{1998b}]{schlegel+98}
{Schlegel}, D.~J., {Finkbeiner}, D.~P., and {Davis}, M.: 1998b,
\newblock {\em \apj} {\bf 500}, 525

\bibitem[\protect\astroncite{{Smith} and {Spinrad}}{1980}]{smithspinrad80}
{Smith}, H.~E. and {Spinrad}, H.: 1980,
\newblock {\em \pasp} {\bf 92}, 553

\bibitem[\protect\astroncite{{Sparks} et~al.}{1994}]{sparks+94}
{Sparks}, W.~B., {Biretta}, J.~A., and {Macchetto}, F.: 1994,
\newblock {\em \apjs} {\bf 90}, 909

\bibitem[\protect\astroncite{{Urry} and {Padovani}}{1995}]{urrypadovani95}
{Urry}, C.~M. and {Padovani}, P.: 1995,
\newblock {\em \pasp} {\bf 107}, 803

\bibitem[\protect\astroncite{{V{\' e}ron-Cetty} and {V{\' e}ron}}{2003}]{VCV03}
{V{\' e}ron-Cetty}, M.-P. and {V{\' e}ron}, P.: 2003,
\newblock {\em \aap} {\bf 412}, 399

\bibitem[\protect\astroncite{{Waters} and {Zepf}}{2005}]{waterszepf05}
{Waters}, C.~Z. and {Zepf}, S.~E.: 2005,
\newblock {\em \apj} {\bf 624}, 656

\bibitem[\protect\astroncite{{Williams} and {Gull}}{1985}]{williamsgull85}
{Williams}, A.~G. and {Gull}, S.~F.: 1985,
\newblock {\em \nat} {\bf 313}, 34

\bibitem[\protect\astroncite{{Wilson} et~al.}{2001}]{wilson+01}
{Wilson}, A.~S., {Young}, A.~J., and {Shopbell}, P.~L.: 2001,
\newblock {\em \apj} {\bf 547}, 740

\end{thebibliography}

\end{document}